\documentclass[aps,prl,reprint,groupedaddress,nofootinbib]{revtex4-1}
\usepackage{hyperref}
\usepackage{amsmath}
 \usepackage{multirow}
\usepackage{array}
\newcolumntype{L}[1]{>{\raggedright\let\newline\\\arraybacksslash\hspace{0pt}}m{#1}}
\newcolumntype{C}[1]{>{\centering\let\newline\\\arraybackslash\hspace{0pt}}m{#1}}
\newcolumntype{R}[1]{>{\raggedleft\let\newline\\\arraybackslash\hspace{0pt}}m{#1}}
\usepackage{float}
\usepackage{graphicx}
\usepackage{epsfig}
\usepackage{psfrag}
\usepackage{color}
\usepackage{slashed}

\usepackage{amsfonts}
\usepackage{amssymb}
\usepackage{tikz}
\usepackage{tikz}
\usetikzlibrary{positioning,arrows}
\usetikzlibrary{decorations.pathmorphing}
\usetikzlibrary{decorations.markings}

\newcommand*{\be}{\begin{equation}}
\newcommand*{\ee}{\end{equation}}
\newcommand*{\bea}{\begin{eqnarray}}
\newcommand*{\eea}{\end{eqnarray}}

\newcommand{\comment}[1]{}

\newcommand{\cref}[1]{Chapter~\ref{c.#1}}

\def\beq{\begin{equation}}
\def\eeq{\end{equation}}
\def\bea{\begin{eqnarray}}
\def\eea{\end{eqnarray}}
\def\ba{\begin{array}}
\def\ea{\end{array}}
\def\bi{\begin{itemize}}
\def\ei{\end{itemize}}
\def\be{\begin{enumerate}}
\def\ee{\end{enumerate}}
\def\bc{\begin{center}}
\def\ec{\end{center}}
\def\bt{\begin{table}}
\def\et{\end{table}}
\def\btb{\begin{tabular}}
\def\etb{\end{tabular}}

\def\lsim{\raise0.3ex\hbox{$\;<$\kern-0.75em\raise-1.1ex\hbox{$\sim\;$}}}
\def\gsim{\raise0.3ex\hbox{$\;>$\kern-0.75em\raise-1.1ex\hbox{$\sim\;$}}}
	
\usepackage{cancel}
 \usepackage[normalem]{ulem}
\usepackage{color}
\def\comment#1{\textcolor{blue}{\large(\it{#1})}}

\def\lapp{\mathrel{\rlap{\raise.5ex\hbox{$<$}}
                    {\lower.5ex\hbox{$\sim$}}}}
\def\gapp{\mathrel{\rlap{\raise.5ex\hbox{$>$}}
                    {\lower.5ex\hbox{$\sim$}}}}
 \preprint{  
 	TIFR/TH/18-xx
 }
\begin{document}

\title{Unearthing the electroweak structure of warped 5D models}

\author{ Abhishek M. Iyer$^1$ and K. Sridhar$^2$}
\affiliation{
$^1$INFN-Sezione di Napoli, Via Cintia, 80126 Napoli, Italia\\
$^2$Department of Theoretical Physics, Tata Institute of Fundamental Research, Homi Bhabha Road, Colaba, Mumbai 400 005, India}

\begin{abstract}
Heavy charged bosons, with masses in the range of a few TeV, are a characteristic of warped extra-dimensional models with bulk gauge
fields. Rendering the latter consistent with electroweak precision
tests typically requires either a deformation of the metric or extension
of the gauge symmetry. We make here the first attempt at finding
empirical discriminants which would tell these models
apart. Demonstrating the power of simple kinematic observables
involving same-sign leptons, we construct simple yet powerful
statistical discriminants.

\end{abstract}

\maketitle

\noindent 
The Randall-Sundrum (RS) Model, as written down originally, 
invokes an extra space dimension $y$ compactified on an $S^1/Z^2$
orbifold of radius $R$ \cite{Randall:1999ee}.  Two branes (UV and IR
respectively) are located at the orbifold fixed points (end-points)
$y=0$ and $y=\pi R \equiv L$. 
 The bulk has a strong $AdS$ curvature, with its magnitude $k$ only somewhat smaller than the Planck scale $M_{Pl}$,
 The solutions to the vacuum
Einstein equations admit a static background with Lorentz invariance
built in, namely the warped metric
\begin{equation}
	ds^2=e^{-2 A(y)}\eta_{\mu\nu}dx^{\mu}dx^{\nu}- dy^2 \ , \quad 
	A(y) = \pm k \vert y \vert \ .
	\label{e1}
\end{equation}

The gauge-hierarchy problem is resolved by choosing $kR\sim 12$ and with an IR localized Higgs, and electroweak scale at around 250 GeV
materialises naturally on account of the warping.

However, the warping also affects all IR-localised fields and,
in particular, mass scales which suppress dangerous
higher-dimensional operators responsible for proton decay or neutrino
masses are also lowered. This spells a disaster for the model. 
A way out
is to conisder only
the Higgs to be localised on or near the IR brane
\cite{Pomarol:1999ad, Gherghetta:2000qt, Grossman:1999ra}.
Collectively known as Bulk RS models, these viable variations yield a
bonus: localising fermions at different positions in the bulk implies
differing overlaps of their profiles with the Higgs field. This then provides 
a natural mechanism for explaining fermion mass and mixing hierarchy and in addition, suppresses dangerous
flavour-changing neutral currents (FCNCs).
 For reviews of
bulk models, see Refs. \cite{Gherghetta:2010cj} and
\cite{Raychaudhuri:2016cj}.

Electroweak precision tests impose very strong
constraints on bulk models. If, for example, only the gauge bosons
propagate in the bulk, then the couplings of their KK modes to the
IR-localised fermions lead to unacceptably large
contributions to $T$ and $S$, thereby resulting in a lower bound of
about 25 TeV on the mass of the first KK mode of the gauge boson. A
six-dimensional generalization of the RS model brings this mass limit
down to about 7-8 TeV \cite{Arun:2015kva,Arun:2016ela,Arun:2016csq},
but it is possible to stick to the original five-dimensional model and
make other modifications to address the electroweak constraints.
Localising the light fermions close to the UV brane, significantly
reduces the constraints from the $S$-parameter. The corrections to the $T$ parameter
can be softened by enlarging the gauge symmetry in the bulk to $SU(3)_c \times SU(2)_L \times SU(2)_R \times U(1)_y$
\cite{Agashe:2003zs, Agashe:2006at}   
Appropriate 
 choice of the fermion representations also helps suppress contributions in the non-oblique $Z \rightarrow b \bar b$
corrections \cite{Davoudiasl:2009cd, Iyer:2015ywa}.

An alternative to enhanced gauge symmetry is to use a deformed metric
near the IR brane \cite{Cabrer:2010si,Cabrer:2011fb}, with the
softening of the singularity at the IR boundary implying that the
Higgs is a bulk scalar field.  The function $A(y)$ in
Eq. \ref{e1} is now modified to
\begin{equation}
	A(y)=k y - \nu^{-2} \, \log(1- y/y_s)
\end{equation}
(the limit $\nu\rightarrow 0$ reverting to the RS geometry).  The UV
brane is still located at $y=0$. The IR brane is, however, located at
$y=y_1$ with the position of the singularity ($y=y_s$) located behind
it at $y_s \simeq y_1 + {\cal O}(k^{-1})$.  To address the hierarchy
one requires $A(y_1)\sim 35$, which fixes the value of $y_1$.  The deformation
causes the Higgs field to be moved further away from the IR brane
whereas the gauge boson KK modes move towards it. The consequent
reduction in the overlap of the Higgs and KK gauge boson modes reduces
the bounds from the $T$ parameter and the mass of first KK gauge boson
mode can now be as small as 1.5 TeV \cite{Iyer:2015ywa}.

In typical Bulk RS models, the gauge boson KK excitations are the
lightest ones. Constituting the most promising probes, searches of KK
excitations of gluons \cite{Agashe:2006hk, Lillie:2007yh,
  Guchait:2007jd, Allanach:2009vz, Iyer:2016yjb}, electroweak gauge
bosons\cite{Agashe:2008jb, Agashe:2007ki} and the Higgs
\cite{Mahmoudi:2016aib, Mahmoudi:2017txo} have been proposed and some
of these executed by the ATLAS \cite{Aad:2015fna} and CMS experiments
\cite{Chatrchyan:2012ku}.

We are interested in distinguishing between the deformed and the
custodial model at the LHC. In the custodial model, the SM fermion
doublets are extended to fields transforming as $(2,2)$. The quark
multiplet thus contains exotic $\chi_{5/3}$ fermions absent in the
deformed model. We look at the production of these states, in
association with a top, from the decay of the first KK-mode of the
$W^\pm$ bosons. Cascade decays of the $\chi_{5/3}$ fermion leads to
two leptons with the same sign. Such final sates are also possible in
deformed scenarios, where a heavy gauge boson decays in to a
vector-like-quark and a top.  However, the two cases are characterized
by different kinematics and we demonstrate that simple kinematic
variables like the azimuthal separation $\Delta \phi$ between the same
sign leptons and $p_T$ combinations of the same sign leptons are
effective in not only suppressing the background but also in
distinguishing between the two scenarios.

The third generation quarks being
 localized close to the IR brane,
the coupling of a KK-$W^{'+}$ to a $t \bar b$ pair is quite similar to
that to a VLQ--SM quark pair. Thus, owing to the larger phase space
available, the KK-$W^{'+}$ would decay primarily into the former
channel, rendering this the discovery mode.  Several dedicated
analyses in this direction have used combinations of different
variables (both kinematic and substructure) to extract the maximum
signal efficiency. Instead, we perform a minimal LHC analysis to set
up the discovery mode and demonstrate a simple set of cuts to achieve
a $\sim3\sigma$ sensitivity for the process $p~p\rightarrow
W^{'+}\rightarrow t\bar b $.  The matrix element for this process is
generated using {\tt{MADGRAPH 6}} \cite{Alwall:2014hca} using the
model files generated by {\tt{FEYNRULES}}.  To maximize discovery
potential hadronic decay of the top is considered.  Generated events
are passed on to {\tt{PYTHIA 8}} \cite{Sjostrand:2014zea} for
showering and hadronization.  For $m_{W'}\sim 2.5$ TeV, the decay
partons are likely to be associated with very high transverse momentum
($p_T$) jets.  The jet reconstruction radius must be such that the
decay products of the top are captured within a cone of radius $R$,
with the opening angle roughly being $\sim 2 m_t / p^t_T$, where
$p^t_T$ is the top transverse momentum. It is clear that a radius $R=0.5$
is sufficient for the kinematics under consideration. Using
{\tt{FASTJET}}, \cite{Cacciari:2011ma} with the
{\tt{Cambridge-Aachen}} \cite{Dokshitzer:1997in} jet clustering
algorithm and require that the jets satisfy $p_T>100$ GeV. The top
candidate is identified among the two leading jets, using the {\tt{HEP
    TOP TAGGER}} \cite{Plehn:2010st} algorithm.   

Post top-identification, we demand that the invariant mass of the
two jet system satisfies $2000<m_{j_0j_1}<3000$ GeV. To estimate the
background, we simulate hard QCD processes in the following kinematic
regime: require the scalar sum of the visible transverse momenta to be
$\sum p^{j}_T>500$ GeV  and the invariant mass of the outgoing
partons to be $\hat m_{\rm jets}>800$ GeV. Taking into account the width of the resonance as well as the mass resolution of the final states, the background is generated with these choices of parameters.
 These values are chosen as the
decay constituents of $W'$ are likely to have a $p_T$ of at least
500-600 GeV each.  This reduces the QCD cross-section to $10^5$
fb. Furthermore, after the selection criteria detailed above are imposed , the QCD background has a
fake rate of $0.1\%$ while the signal has an acceptance of
$13\%$. Given the current bounds from LHC, the production
cross-section for the $W'$ is chosen to be $20$ fb. \footnote {This is
  the typical value for the production cross-section for charged
  bosons in extra-dimensional models.} This leads to a rough
sensitivity of $S/\sqrt{B}\sim 3$ with $\sim 400~fb^{-1}$ of data. It
is to be stressed that this acceptance can be significantly improved
with even lesser luminosity. However, it has been considered
extensively and is not the goal of this paper. The objective behind
this section was to demonstrate that even the most basic cuts is
sufficient to obtain a reasonable signal acceptance. Post this
discovery, it is then necessary to extract the origin of the heavy
charged object: whether the origin of the heavy gauge state is due to
a SM gauge symmetry or due to and extended gauge group.
One simple way is to look at the charge of the heavy fermion the $W' $
decays into, as determined by analysing the multileptonic final state.

 For custodial models, the electroweak gauge group is extended to
 $SU(2)_L\times SU(2)_R\times U(1)_{B-L}$. The $(t,b)$ doublet is
 replaced by a bidoublet represented by
$Q_3\equiv \begin{bmatrix}
\chi_{5/3}&t\\
T_{2/3}&b
\end{bmatrix}$
where $\chi,T$ are exotic fermions with electro-magnetic charges
$5/3,2/3$ respectively. The crucial difference between the deformed
and the custodial models is the presence of the charge $5/3$ state.
For simiplicity we consider $t_R$ to be a gauge singlet.
	  and consider the
          decay of the $W^{(1)}$ is into $\chi_{5/3}\bar t$. The
          $\chi_{5/3}$ can further decay into $Wt$ resulting in a
          $Wt\bar t$ final state. Considering a total leptonic final
          state, this leads to three leptons with two leptons of the same sign.
	
	

For deformed
  models, the gauge structure is SM like and hence the heavy KK
fermions also have charges $Q =2/3,-1/3$. 
The aforementioned three lepton final state can arise in two ways:
A)	 The VLQ decays into a $tZ$ or $bZ$ as the case may
          be. Thus, the net final state from the gauge KK state is
          $t\bar bZ$. Assuming both the top and the $Z$ decay
          leptonically, we have a 3 lepton final state. with 2 leptons
          of same sign.  This case can however be distinguished using
          a $Z$ mass veto for two leptons with opposite sign.\\
	B)	 If the KK fermion decays into $W \bar b$ ($Q=2/3$) and
          $W^- t$ ($Q=-1/3$), the overall final state from the
          gauge-KK would be $W\bar b b$ and $W\bar t t$
          respectively. The former leads to only a single isolated
          hard lepton, while the latter may lead to three leptons and
          two b-tagged jets.


\subsection{Variables for discrimination:} In order to distinguish the two scenarios, it is necessary to understand the kinematic features for the two gauge structures. Since the signal is characterized by the presence of two leptons of the same sign, it is useful to construct variables using these two leptons:\\

\begin{figure}[htb!]
\begin{center}
\begin{tabular}{cc}
\includegraphics[width=4.2cm, height=3cm]{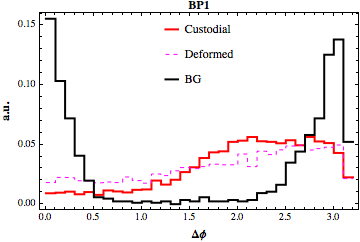}&
\includegraphics[width=4.2cm, height=3cm]{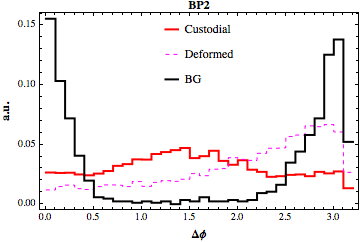}\\
\includegraphics[width=4.2cm, height=3cm]{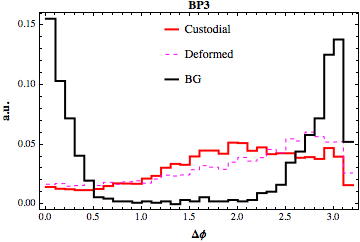}&
\includegraphics[width=4.2cm, height=3cm]{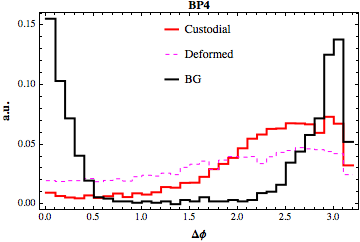}
\end{tabular}
\end{center}
\caption{ $\Delta\phi$ between the same sign leptons for the 4 benchmark points.}
\protect\label{delphi}
\end{figure}

\textbf{$\Delta\phi_{l^\pm l^\pm}$ between the same-sign leptons:} For
the custodial case, the two same sign leptons originate from the decay
of the $5/3$ state. As a result the $\Delta\phi$ between them would
depend on the boost of the VLQ. For the deformed case, on the other
hand, one of them is due to the decay of the VLQ while the other is
due to the top originating from the heavy charged vector boson. Since the
VLQ and the top from the charged boson are produced back-to-back,
 the same-sign leptons from them are also broadly separated. As a
result the utility of this variable to segregate the two cases depends
on the benchmark point used. We consider the following four different combinations of bench mark points:\\
\begin{eqnarray}
\text{BP1}::(3000,1500)\;\;\text{BP2}::(3000,1000)\nonumber\\\text{BP3}::(3500,1500)\;\;\text{BP4}::(3500,2000)
\label{bp}
\end{eqnarray}
where are masses are expressed as $(m_{W'},m_{VLQ})$.
 For the background the same sign lepton
is due to radiation off one of the tops and hence has a very distinct
distributions where the $\Delta\phi_{l^\pm l^\pm}$ are either back-to-
back or in the same direction. This is extremely useful in
distinguishing the background from the signal. Fig. \ref{delphi} gives
the distribution of this variables for the different benchmark points
and the background. The deformed models are characterized by a fairly
similar distribution for all the benchmark points.  It can be seen
that the efficiency of this variable is better for BP2 over others and
can be attributed to the larger boost of 1 TeV VLQ states from the
decay of $W'$. \\

\begin{figure}[htb!]
\begin{center}
\includegraphics[width=4.2cm,height=3.5cm]{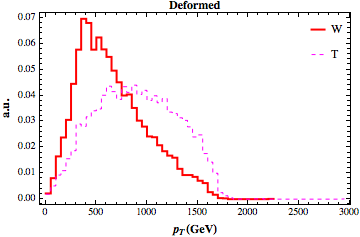}	
\includegraphics[width=4.2cm,height=3.5cm]{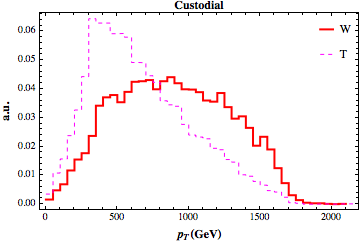}
\includegraphics[width=4.2cm,height=3.5cm]{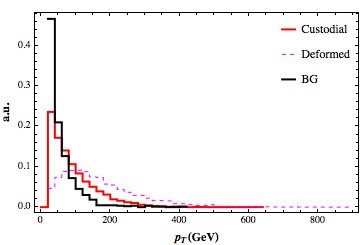}
\end{center}
	\caption{ Plots in the top row give the parton $p_T$ of the decay products of the VLQ  for deformed (left) and for custodial (right) scenarios. The bottom plot gives the distribution of the minimum transverse momentum between the same sign lepton combination $p^{min}_T=min(p_T^{l_1^\pm},p_T^{l_2^\pm})$ }
	\protect\label{pt}
\end{figure}

\textbf{$p^{min}_T$ of the same sign-lepton combination ($p^{min}_T=min(p_T^{l_1^\pm},p_T^{l_2^\pm})$).}
For the deformed models one of the same sign leptons is due to the SM $W$ originating from the decay of the VLQ while the other is due to the top from the heavy gauge boson vertex. Due to the boost of the top, the corresponding lepton from this top is characterized by larger $p_T$ than the one due the $W$ from the decay of the VLQ.  Left plot of Fig. \ref{pt} gives the comparison of parton level $p_T$ for the $W$ and the top from the VLQ. While the $W$ is characterized by smaller transverse momentum, it is only shared between lepton and the neutrino. Resultantly, the $p^{min}_T$ of the same sign lepton combination is likely to have a momentum distribution peaking at relatively larger values as shown by the dashed-pink line at the bottom plot of Fig. \ref{pt}.

For custodial models on the other hand, the two same-sign leptons are from the decay products of the VLQ. The right plot of Fig. \ref{pt} gives a comparison of the $p_T$ for $W$ and the top from $\chi_{5/3}$ VLQ.  Since the $p_T$ is shared between three objects: $b$-jet, lepton and neutrino, the corresponding lepton is characterized by relatively lower $p_T$ than in the case when the decay proceeds due a non-custodial scenario. This is evident by the distribution of the solid red line in the bottom plot in Fig. \ref{pt}.

Given the two distributions, it is necessary to develop a quantitative measure to distinguish them. 
We assume that one
particular Hypothesis say $H_T$, to be true, which is to be tested against the alternative hypothesis say $H_A$.
To estimate the number of events \textit{N} required to disfavour a given spin hypothesis $H_A$ to some factor $R$, we solve
\begin{equation}
\frac{1}{R} = \frac{p (H_A| N \text{~events from~}H_T)}{p(H_T | N
	\text{~events from~}H_T)} \label{ratio}
\end{equation}
where $R$ is integer and implies that the alternative hypothesis $H_A$ is disfavoured at $1:R$ odds in favour of $H_T$. Following \cite{Athanasiou:2006ef,Allanach:2017qbs}, we present results for $R=20,1000$.
The \textit{N} events are characterized by their values of either one or a set of observables $\mathcal{O}_i$. In the first instance we choose two possibilities: either  $\mathcal{O}_1=\Delta \phi$ or $\mathcal{O}_1= p^{min}_T$ between the same-sign lepton candidates. 
We find that while the former is useful for background discrimination, the latter is more efficient for distinguishing the two models. Following, this we adapt an analysis involving both the variables simultaneously. Following the steps in \cite{Allanach:2017qbs} for the discrete implementation of  
Kullback-Leibler divergence \cite{Athanasiou:2006ef}, Eq. \ref{ratio} becomes: 


%
\begin{equation}
\log\left( \frac{1}{R}\right) =
\sum_{i=1}^K \left[\mu ^{(T)}_j \log \frac{\mu_j^{(A)}}{\mu_j^{(T)}} +
\mu_j^{(T)} - \mu_j^{(A)}\right]. \label{penultimate} 
\end{equation}
where $\mu_j$ is the expectation value for the number of events in the $j^{th}$ bin for a given hypothesis.
To translate the above expression into the number of events (and hence the integrate luminosity $\mathcal{L}$) required to separate $H_A$ from $H_T$ at $1:R$ 
we use
\begin{equation}\mu_j^{(X)}=\mathcal L~\sigma_{tot}^{(W')}~\text{B.R}(W'\rightarrow l^+l^-l^++ X)
\epsilon^{(X)}_j, \label{mui}
\end{equation}
where $X$ denotes jets and missing energy and $\epsilon_j$ is the collider acceptance efficiency for the $j^{th}$ bin.
\footnote{We assume  that $\sigma_{tot}^{(W')}$ is same for both the hypothesis $X=A,T$.}

Using Eq.~\ref{penultimate} and Eq.~\ref{mui} and ${
	N}_R=\mathcal L \sigma_{tot}$, the number of events $N_R$ of the true hypothesis $H_T$ to disfavour $H_A$ at $1:R$ odds is
\begin{equation}
N_R = \frac{\log  R}{\sum_{j=1}^K
	\left[\epsilon ^{(T)}_j \log \frac{\epsilon_j^{(T)}}{\epsilon_j^{(A)}} + 
	\epsilon_j^{(A)} - \epsilon_j^{(T)}\right]}.
\label{lr}
\end{equation}
\textbf{Results of the analysis:} We employ this analysis for the different benchmark masses in Eq. \ref{bp}.
The typical model cross-section for 3 TeV state for deformed models is $<1$ fb while that for custodial models $\sim 15$ fb for deep UV localization of the light quarks.  One may argue that on account of the different cross-sections a relatively `early' discovery is more likely to be a sign of custodial models over deformed models. However,
direct searches in the di-jet ($tb$) final state may give a clear hint of the underlying resonance only if it has a sufficiently narrow width.
 In the event of a broad-width, the mass resolution is not likely to be as precise. For the purpose of comparison we assume a similar production cross-section for both scenarios. \footnote{The production cross-sections for the deformed case can be enhanced by slightly enhancing the couplings of the light quarks to the KK gauge fields while respecting a $U(2)$ symmetry in the coupling space.}. Upper limits exist on $\sigma(pp\rightarrow W')\times B.R.(W'\rightarrow tb)$ from direct searches on the $tb$ final state \cite{Aaboud:2018juj} where masses below 3 TeV with $\sigma(pp\rightarrow W')\times B.R.(W'\rightarrow tb) >15$ fb are excluded.
We assume a production cross section of $15~fb$ for 3 TeV and 5 fb for $3.5$ TeV. Table  \ref{tab2} gives the results of the statistical discussion using using both $p_T^{min}-\Delta\phi_{l^\pm l^\pm}$. We present results for both $R=20$ (black) and $R=1000$ (red).
 The probabilities are computed by constructing bins of sizes $(0.7,145)$ in the $\Delta\phi_{l^\pm l^\pm}-p_T^{min}$ over the range $[0-\pi,0-570]]$.
While $p_T^{min}$ of the same sign lepton combination is extremely useful in distinguishing the two scenarios, it is not as effective as $\Delta\phi_{l^\pm l^\pm}$ for background discrimination. 
  Note that the  conclusions  using both variables are expected to be similar within statistical fluctuations  to those  which takes only $p_T^{min}$ into account. This is because $p_T^{min}$ plays the dominant role for the discrimination in both cases while $\Delta\phi_{l^\pm l^\pm}$ is practically a dummy variable as far as the discrimination between the two signal possibilities are concerned. We reiterate  that the role  $\Delta\phi_{l^\pm l^\pm}$ is primarily restricted to segregating both the signal possibilities from the background. Given the drastically different distributions of $\Delta\phi_{l^\pm l^\pm}$ for the background from the signal possibilities 3-4 events are suffice to eliminate the background only hypothesis at a $1:20$ odds.

\begin{table}[htb!]
	\begin{center}
		\begin{tabular}{|ccc|ccc|} 
			\hline
			\multicolumn{3}{| c| }{BP1}&\multicolumn{3}{ c| }{BP2}\\
			$N_R$           & Custodial  & Deformed&$N_R$           & Custodial  & Deformed  \\ \hline

			Custodial         &$\infty$&\begin{tabular}{c}16 \\\textcolor{red}{39}\end{tabular}&Custodial&$\infty$&\begin{tabular}{c}13 \\\textcolor{red}{30}\end{tabular}\\

			Deformed     &\begin{tabular}{c}11 \\\textcolor{red}{27}\end{tabular}&$\infty$&Deformed&\begin{tabular}{c}6 \\\textcolor{red}{15}\end{tabular}&$\infty$\\ \hline  \hline
			\multicolumn{3}{| c| }{BP3}&\multicolumn{3}{ c| }{BP4}\\
			$N_R$           & Custodial  & Deformed&$N_R$           & Custodial  & Deformed  \\ \hline

			Custodial         &$\infty$&\begin{tabular}{c}20 \\\textcolor{red}{47}\end{tabular}&Custodial&$\infty$&\begin{tabular}{c}14 \\\textcolor{red}{34}\end{tabular}\\

			Deformed     &\begin{tabular}{c}12 \\\textcolor{red}{29}\end{tabular}&$\infty$&Deformed&\begin{tabular}{c}9 \\\textcolor{red}{22}\end{tabular}&$\infty$\\ \hline
			
		\end{tabular}
		\caption{\label{tab2} Table gives the expected number of events $N_R = {\mathcal L}
			\sigma_{tot}^{(X)}$, to disfavour the column hypothesis ($H_A$) in favour of the row hypothesis ($H_T$)
			by a factor of $R=20$ (black) and $R=1000$ (red) at the 13
			TeV 
			LHC. Both $p^{min}_T$ and $\Delta\phi_{l^\pm l^\pm}$ variables are used in this case.} 
	\end{center}
\end{table}

\textbf{Discussions:}
The results in Table \ref{tab2} can be be converted to the required luminosity by simply assuming a production cross-section for the $W'$ and the branching fraction for the VLQ into the corresponding states. For simplicity we discuss the results for $R=20$ and assume equal production cross-sections and branching fractions for both the models.
If we assume a production cross-section of $15~fb$, with an integrated luminosity of $3000$~fb$^{-1}$, branching fractions as low as $20\%$ for the VLQ $T$ \footnote{We assume $B.R.(W'\rightarrow T X)\sim 50\%$, where X is SM state.} can be probed.
 The results of Table \ref{tab2} are very general and can be used to discriminate other classes of models with a similar gauge and fermion content.

Heavy charged gauge bosons are a characteristic feature of several extensions beyond the Standard Model. 
 Corresponding to the gauge origins of these heavy vectors, the representation of the fermion content also differ. Using this as a motivation, we present a  methodology to distinguish the two cases in the event of a discovery. 
 We demonstrate  effectiveness of this technique by using a  statistical tool which utilities
  simple kinematic variables like $\Delta \phi$ between the same-sign leptons and $p_T$.
 Given the generic nature of the method, the analysis and the corresponding  results can also be extended with similar gauge and fermion content.
\section*{Acknowledgments}
We are grateful to Debajyoti Choudhury for  several useful discussions and a careful reading of the manuscript.
A.I. was supported in part by MIUR under Project No. 2015P5SBHT and by the INFN research initiative ENP. K.S. 
work is partly supported by a project grant from the Indo-French Centre for the Promotion of Advanced Research 
(project no. 5904-2).

\bibliography{mybib1_new}
\end{document}